\documentclass[epjST]{svjour}
\usepackage{graphics}
\usepackage{color}
\newcommand{\beq}{\begin{equation}}
\newcommand{\eeq}{\end{equation}}
\newcommand{\bea}{\begin{eqnarray}}
\newcommand{\eea}{\end{eqnarray}}
\def\lsim{\buildrel < \over {_{\sim}}}
\def\gsim{\buildrel > \over {_{\sim}}}
\begin{document}
\title{Unraveling the Flux-Averaged Neutrino-Nucleus Cross Section}
\subtitle{}
\author{Omar Benhar\thanks{\email{omar.benhar@roma1.infn.it}}}
\institute{INFN and Department of Physics, Sapienza University, I-00185 Rome, Italy}
\abstract{
The interpretation of the nuclear cross sections measured using accelerator neutrino beams involve 
severe difficulties, arising primarily from the average over the incoming neutrino flux. 
The broad energy distribution of the beam particles hampers the determination of the energy transfer 
to the nuclear target, the knowledge of which is needed to pin down the dominant 
reaction mechanism. Overcoming this problem requires the development of a theoretical approach suitable
to describe neutrino interactions at energies ranging from hundreds of MeV to few GeV. 
In this paper, it is argued that the approach based on the factorisation of the nuclear cross section provides a 
consistent framework for the calculation of neutrino-nucleus interactions in both the quasi elastic and
inelastic channels. The near-degeneracy between theoretical models based on different assumptions, and the 
use of electron scattering data to advance the understanding of neutrino-nucleus cross sections are also discussed.
}
\maketitle
\section{Introduction}
\label{intro}
{\em Varius, multiplex, multiformis}.  The three latin adjectives, famously employed to portray the complex personality of roman emperor Hadrian \cite{memoirs} provide a remarkably accurate characterisation of the nuclear response to neutrino interactions. 
Owing to the broad energy spectrum of the beam particles, the effects of the many and diverse mechanisms contributing to the 
neutrino-nucleus cross section are, in fact, inextricably tangled with one another, and the interpretation of the detected signals involves
daunting challenges. A comprehensive and clear discussion of the main issues of the physics of neutrino interactions can be found
Ref.~\cite{singh:book}.

The interest in theoretical modelling of neutrino-nucleus scattering was suddenly ignited in the early 2000s, when
the treatment of nuclear effects\textemdash until then largely ignored\textemdash was clearly recognised as a major source of systematic error in accelerator-based
searches of neutrino oscillations~\cite{NUINT}. Few years later, the inadequacy of the Relativistic Fermi Gas Model (RFMG)\textemdash commonly employed in Monte Carlo simulations of neutrino interactions\textemdash was clearly exposed by its conspicuous failure to explain the flux-averaged double-differential  $\nu_\mu$-carbon cross section measured by the MiniBooNE Collaboration \cite{CCQE}.

Over the past few years a number of experimental studies have been devoted to the measurement of neutrino cross sections
using a variety of targets~\cite{MINERVA,T2K,MICRO,NOVA}. The latest data have been recently reported 
at the Workshop on New Directions in Neutrino-Nucleus Scattering, held at FNAL on March 15-18, 2021~\cite{prenuint}.

From the observational point of view, neutrino interaction 
events are classified according to the number of detected pions. Zero- and one-pion events\textemdash referred to as $0\pi$ and 
$1\pi$\textemdash are mainly associated with quasi elastic scattering and resonance production, respectively, while two or more
pions are believed to be produced in deep inelastic processes.
The understanding of this body of data requires a consistent model  of the 
neutrino-nucleus cross section, suitable for use in a broad range of target masses and kinematical conditions.

The studies carried out over the past two decades have led to the development of a number of advanced
models of quasi elastic neutrino-nucleus scattering, taking into account
both the effects of strong interaction dynamics and the variety of mechanisms contributing to the flux-averaged cross section~\cite{benhar_PRD,coletti,ankowski,rocco,martini,nieves,SuSa,GIBBU,natalie}.
Some of these models have arguably reached the
degree of maturity required for a meaningful comparison between their predictions and the available data.
In this context, an essential role is played by the availability of a large database of electron-nucleus
scattering cross sections, precisely measured using a variety of targets and spanning a broad kinematical 
range, see Ref.~\cite{Benhar:2006wy}. 

Although the description of the measured neutrino-nucleus cross sections
involves non trivial additional problems\textemdash arising mainly from the flux average~\cite{benhar_nufact11} and the treatment of the axial-vector 
contributions to the current driving neutrino-nucleon interactions\textemdash the ability to explain electron-nucleus scattering data, in which 
the relevant reaction mechanisms can be unambiguously identified, must obviously be regarded as a requisite to be met by any models of the nuclear response to weak interactions.
The theoretical approach based on factorisation of the nuclear cross section and the Green's function formalism\textemdash which allows to combine an accurate description of the nuclear target with a fully relativistic treatment of the elementary interaction 
vertex\textemdash has been very successful in explaining electron-nucleus data. Over the past decade, the factorisation {\em ansatz}, 
whose extension to the treatment of weak interactions  
does not involve any conceptual difficulties, has established itself as a viable scheme to study the flux-integrated neutrino-nucleus cross section. 

In many instances, the predictions of theoretical models of neutrino-nucleus interactions turn out to be in satisfactory agreement with experiments. However,  a deeper scrutiny reveals a puzzling feature: models based on conceptually different\textemdash and sometimes even contradictory\textemdash assumptions yield similar results. In view of applications to the different kinematical regimes and nuclear targets relevant to future experiments, the sources of degeneracy between different theoretical approaches\textemdash implying that the agreement between theory and data may in fact be accidental\textemdash need to be identified, so as to firmly assess their predictive power.

Considerable progress has been also achieved by the {\em ab initio} approach based on Quantum Monte Carlo techniques,  which allows to carry out very accurate calculations of the electromagnetic and weak responses of
nuclei as heavy as carbon~\cite{GFMC1}. This approach, however, while providing a
remarkably good account of inclusive electron-nucleus scattering data, is inherently limited  to the kinematical regime in which the non relativistic approximation is expected to be applicable~\cite{GFMC2}. A pioneering application of the GFMC technique to the analysis 
of MiniBooNE and T2K data is discussed in~\cite{PRX}.

The structure of the neutrino-nucleus cross section and its role in the analysis of the observed event rate distributions are outlined in Sect.~\ref{sec2}.
Section~\ref{QE_degeneracy} is devoted to a discussion of the mechanisms contributing to the cross section in the quasi elastic channel, with an emphasis on the uncertainties implied in both the theoretical description and the interpretation of the data. The treatment of quasi elastic one- and two-nucleon emission 
processes based on factorisation of the nuclear transition amplitudes is analysed in Sects.~\ref{singlenucleon} and \ref{extended}. The extension of the this approach to the study of
inelastic processes\textemdash in the broad kinematical region relevant to resonance production and deep inelastic scattering\textemdash  is reviewed in Sect.~\ref{inelastic}. Finally, Sec.~\ref{sec4} summarises the achievements and unresolved issues of  theoretical studies of neutrino-nucleus interactions, and lays down the
prospects for future developments.

\section{Event rate distribution and neutrino-nucleus cross section}
\label{sec2}

Neutrino experiments measure the event rate distribution as a function of the visible energy, defined as~\cite{physrep:nu}
\begin{eqnarray}
\label{eq:rate}
R^\alpha_\beta(E_\mathrm{vis})=
N\int dE\,\Phi_\alpha(E)\,\sigma_\beta(E,E_\mathrm{vis})\,\epsilon_\beta(E)\,P(\nu_\alpha\rightarrow\nu_\beta,E) \ ,
\end{eqnarray}
where $N$ is a normalisation factor, the indices $\alpha$ and $\beta$ specify the neutrino flavour, $\Phi_\alpha(E)$ and $E$ are 
the neutrino flux and the true neutrino energy, respectively, and $P(\nu_\alpha\rightarrow\nu_\beta,E)$ is the oscillation probability.
 The differential cross section $\sigma_\beta(E,E_\mathrm{vis})$
describes the probability that a neutrino of
energy $E$ produces a distribution of visible energies $E_\mathrm{vis}$ in the
detector, while $\epsilon_\beta(E)$ denotes the detection efficiency.

\begin{figure}[h!]
\begin{center}
\resizebox{0.495\columnwidth}{!}{\includegraphics{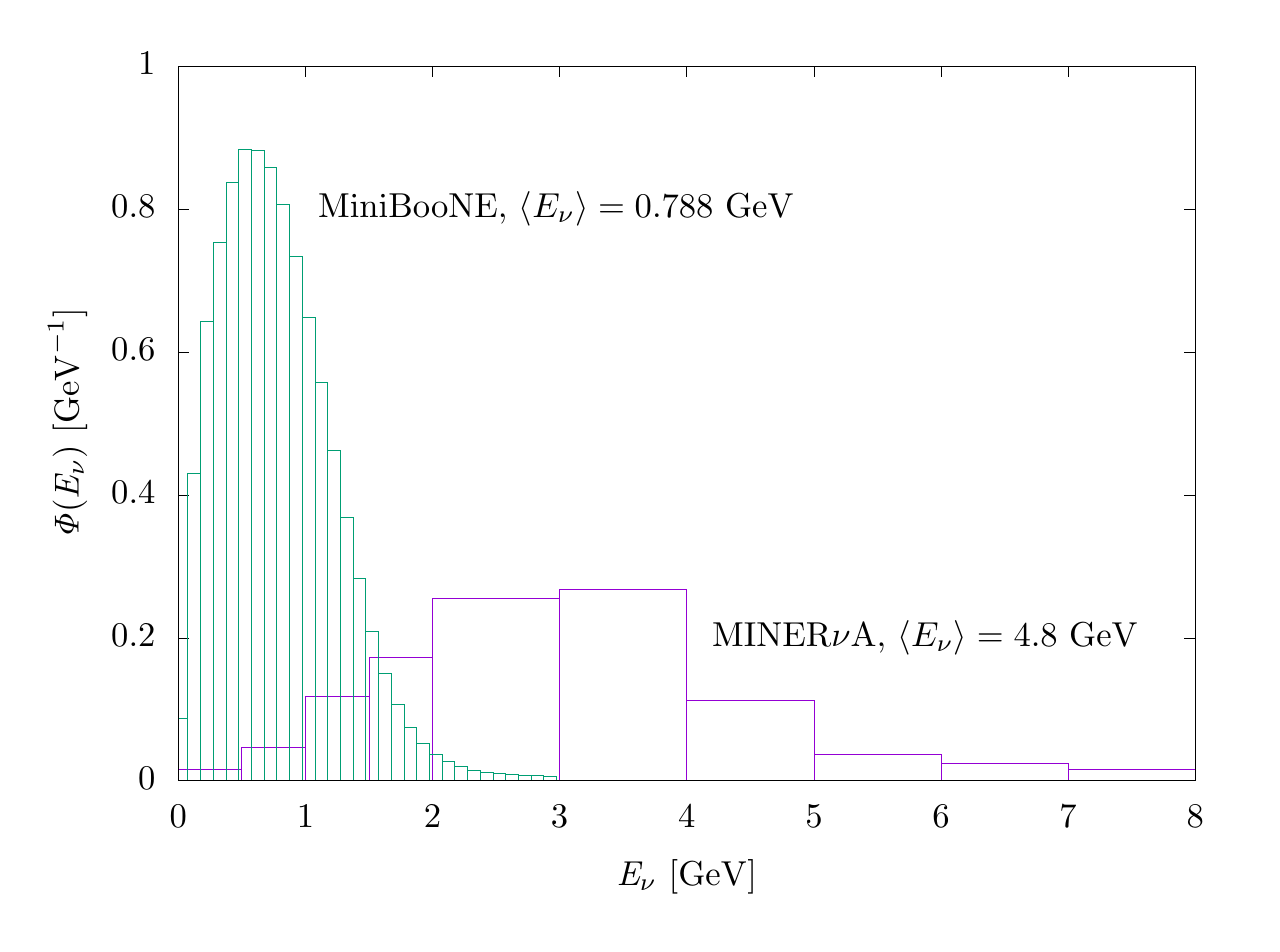} }
\resizebox{0.475\columnwidth}{!}{\includegraphics{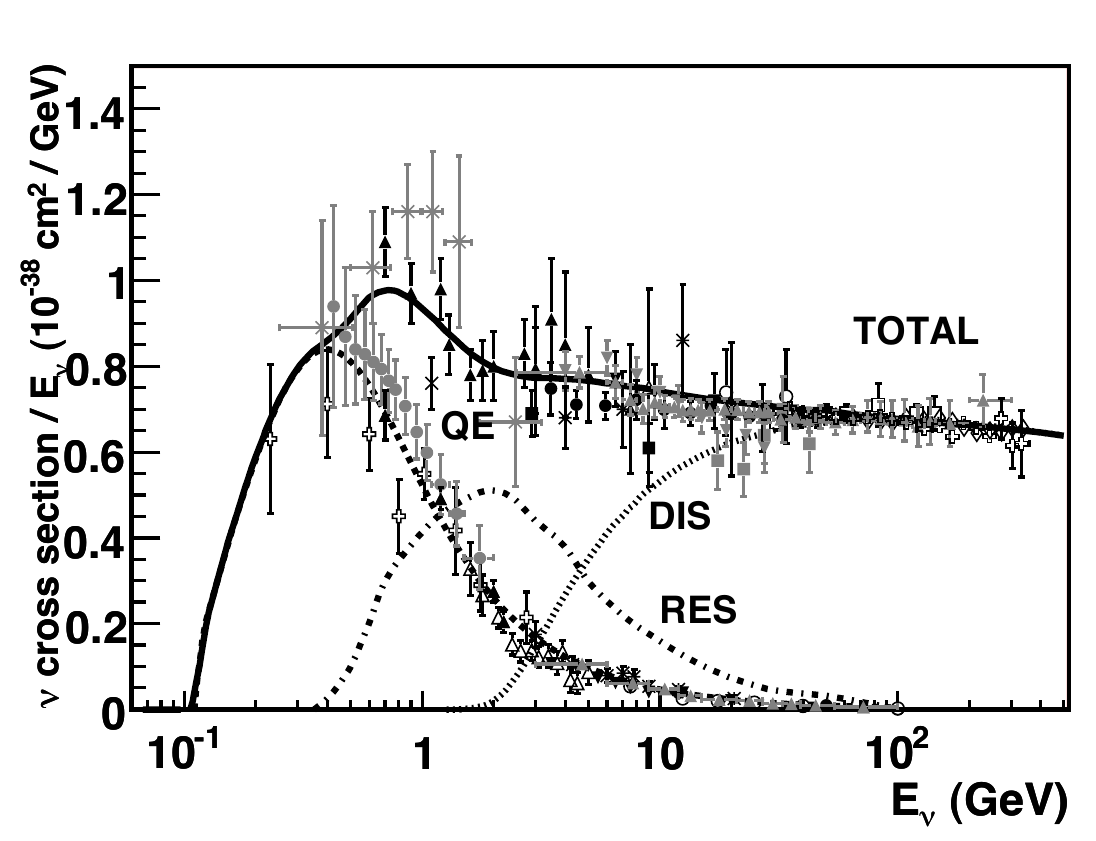} }
\end{center}
\caption{Left: unit-normalised neutrino fluxes incident on the MiniBooNE~\cite{CCQE} and MINER$\nu$A~\cite{MINERVA_flux} detectors.
 Right: contributions of charged current quasi elastic scattering (QE), resonance production (RES) and
deep-inelastic scattering (DIS)  to the neutrino-nucleon cross section, divided by the beam energy. The thick solid
line represents the sum of all contributions~\cite{zeller_RMP}}
\label{complexity}       
\end{figure}

The difficulties involved in the interpretation of the observed event rate can be gauged from Fig.~\ref{complexity}. In the left panel, the 
unit-normalised neutrino fluxes incident on the MiniBooNE~\cite{CCQE} and MINER$\nu$A~\cite{MINERVA_flux} detectors,  are displayed as a function 
of neutrino energy, while the right panel illustrates the contributions of charged current quasi elastic (QE) scattering, resonance production (RES) and
deep-inelastic scattering (DIS)  to the neutrino-nucleon cross section~\cite{zeller_RMP}. It is apparent that all reaction mechanisms
contribute significantly to the flux-averaged signal, and different contributions largely overlap. As a consequence, the determination of $\sigma_\beta(E,E_\mathrm{vis})$, which is critical to 
the oscillation analysis, requires a quantitative understanding of the corresponding cross sections.

The differential cross section of the neutrino-nucleus scattering process
\beq
\nu_\mu + A \to \mu^- + X \   ,
\eeq
where $A$ and $X$ denote the target nucleus of mass number $A$ 
and the hadronic final state, respectively,
can be schematically written in the form
\beq
\label{dsigmaA:1}
{d \sigma}_A  \propto  L_{\alpha \beta} W_A^{\alpha \beta} \ ,
\eeq
where the tensor $L_{\alpha \beta}$ is fully specified by the lepton kinematical variables, while the target response tensor
\beq
\label{dsigmaA:2}
W_A^{\alpha \beta} = \sum_X  \left[ \langle0 |  {J^\alpha_A}^\dagger  | X \rangle
                                                            \langle X |  {J^\beta_A} | 0 \rangle + {\rm h.c.} \right] 
                                                             \delta^{(4)}(P_0 + q - P_X)  \ , 
\eeq
contains all the information on nuclear structure and dynamics. The above equation shows that
the description of the nuclear response involves the target initial and final states, carrying four-momenta $P_0$ and $P_X$, as well as the nuclear current operator
\beq
\label{nuclear:current}
 {J^\alpha_A} = \sum_i {j^\alpha_i} + \sum_{j>i} {j^\alpha_{ij}} \ ,
\eeq
comprising one- and two-nucleon terms.
The sum in Eq.~(\ref{dsigmaA:2}) includes contributions from all possible final states, excited through different reaction mechanisms whose relative weight
depends on kinematics.

\section{QE events and the degeneracy issue}
\label{QE_degeneracy}

Charged Current Quasi elastic (QE) scattering off an individual nucleon, that is, the process corresponding to the final state
\beq
\label{1p1h}
| X \rangle = | p, (A-1)  \rangle \ ,
\eeq
is the dominant mechanism in the kinematical region relevant to the analysis of, e.g.,  the MiniBooNE data, collected using
a neutrino flux of mean energy $\langle E_\nu~\rangle~=~880 \ {\rm MeV}$, see Fig.~\ref{complexity}. 

From the experimental point of view, QE processes 
are characterised by the absence of pions in the final state, and are therefore classified as $0\pi$ events. They are 
fully specified
by the measured kinetic energy and emission angle of the muon, with the knocked out proton and the recoiling nucleus being undetected.
Note that the spectator $(A-1)$-nucleon system can either be in a bound state
or include a nucleon excited to the continuum\footnote{Theoretical studies of the momentum distribution sum rule in isospin-symmetric nuclear matter strongly suggest that
the contribution of $(A-1)$-nucleon states involving more than one particle in the continuum is negligibly small~\cite{BFF}.}.
For example, in the case of a carbon target the state of the recoiling
system can be $| ^{11}{\rm C}^*  \rangle$,  $| p , {^{10}{\rm B}^*}   \rangle$ or $|  n,  {^{10}{\rm C}^* }  \rangle$, where the asterisk indicates that 
the nucleus can be found in any bound states. The corresponding $A$-nucleon final states are
\beq
\label{1p1h_C}
| X \rangle = | p,  {^{11}{\rm C}^*}  \rangle  \ ,
\eeq
or
\beq
\label{2p2h_C}
| X \rangle = | p p,  {^{10}{\rm B}^*}  \rangle \ \ , \ \  | p n,  {^{10}{\rm C}^*}  \rangle  \ .
\eeq
The states appearing in the righ-hand side of Eqs.~(\ref{1p1h_C}) and  (\ref{2p2h_C}) are referred to as
one-particle\textendash one-hole (1p1h) and two-particle\textendash two-hole (2p2h) states, respectively.

The appearance of 2p2h final states in scattering processes in which the beam particle couples to an individual nucleon originates
from nucleon-nucleon correlations in the target ground state or final state interactions (FSI) between the struck particle and the spectator nucleons.
These mechanisms are not taken into account by models based on the independent particle picture of the nucleus, such as the RFGM, according to
which single nucleon knock out  can only lead to transitions to 1p1h final states. On the other hand, transitions to 2p2h states are always allowed in processes driven by two-nucleon
meson-exchange currents (MEC), see Eq.(\ref{nuclear:current}), such as those
in which the beam particle couples to a $\pi$-meson exchanged between two interacting nucleons.
A detailed discussion of the contributions of 1p1h and 2p2h final states to the nuclear response can be found in Refs.~\cite{BLR,RLB,EFS}.
Obviously, the amplitudes of processes involving one- and two-nucleon currents and the same 2p2h final state 
contribute to the nuclear cross section both individually and through interference.

More complex final states, that can be written as a superposition of 1p1h states according to
\beq
\label{RPA_C}
| X \rangle = \sum_n C_n | p_n h_n \rangle  \ ,
\eeq
appear in processes in which the momentum transfer is shared between many nucleons.
The contribution of these processes is often described within the Random Phase Approximation  (RPA),
which amounts to taking into account the so-called ring diagrams to all orders, using phenomenological effective interactions to describe nuclear dynamics~ \cite{RPA1,RPA2}.
On the basis of very general quantum-mechanical considerations,  long-range correlations associated with the final states of Eq.~(\ref{RPA_C}) are expected to become
important in the kinematical region in which the space resolution of the beam particle is much larger than the average
nucleon-nucleon distance in the nuclear target, $d$, i.e. for typical momentum transfers $|{\bf q}| \ll  \pi/d \sim 400  \ {\rm MeV}$.

\begin{figure}[h!]
\begin{center}
\resizebox{0.5\columnwidth}{!}{\includegraphics{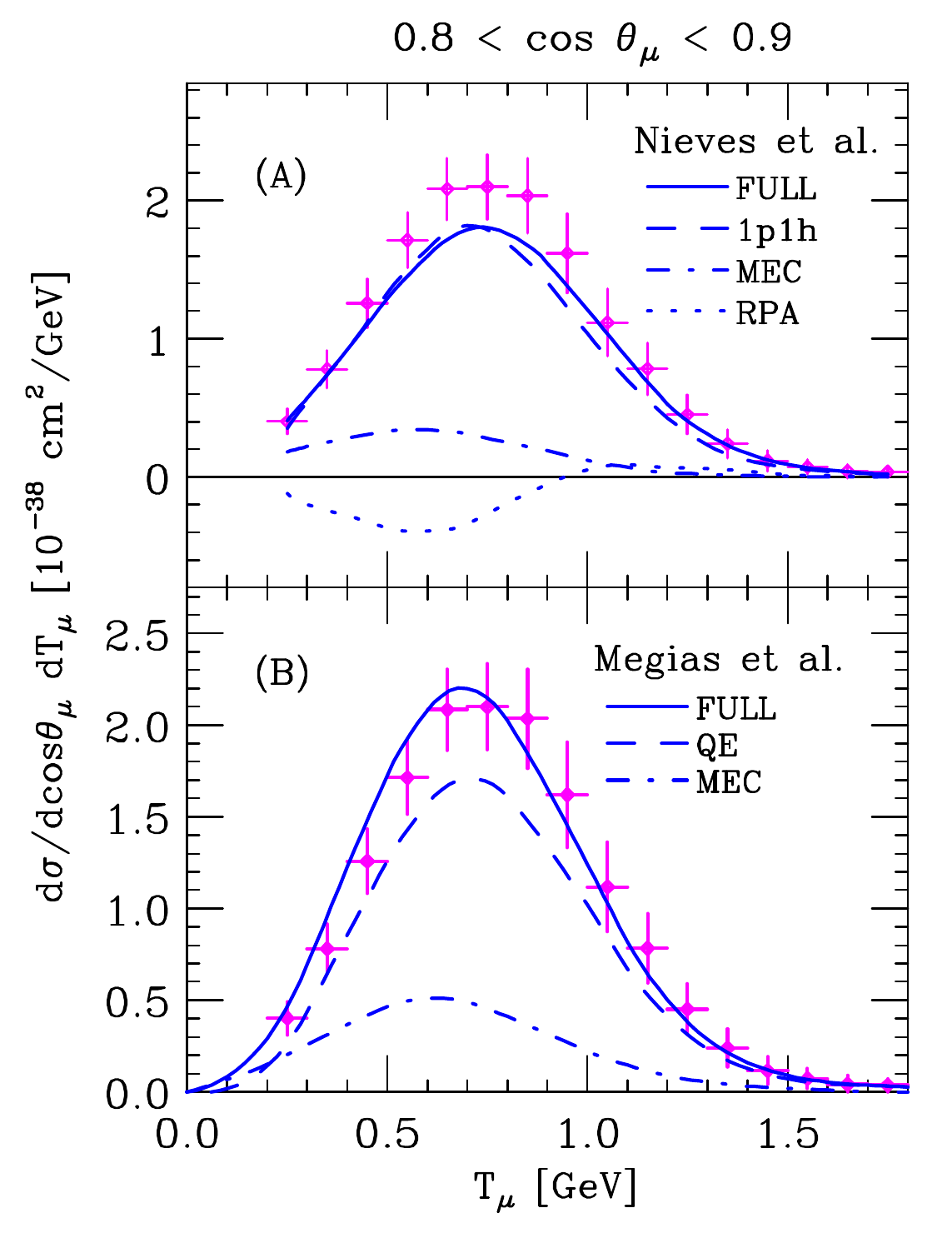} }
\end{center}
\caption{Comparison between the flux-integrated
double differential $\nu_\mu$-carbon cross section in the QE channel measured by the
MiniBooNE Collaboration~\cite{CCQE} and the
results obtained from the models of Nieves {\em et al.}~\cite{nieves} (A), and Megias {\em et al.}~\cite{SuSa}  (B). The solid lines
correspond to the full calculations. The meaning of the dashed, dot-dash and dotted lines is explained
in the text.}
\label{degeneracy}      
\end{figure}

The role played by the reaction mechanisms taken into account by two different models of neutino-nucleus interactions is illustrated in Fig.~\ref{degeneracy},
showing a comparison between the flux integrated double-differential QE cross section measured by the MiniBooNE Collaboration~\cite{CCQE}
and the theoretical results reported in Refs.~\cite{nieves} [panel (A)] and \cite{SuSa} [panel (B)].

Within the approach of Ref.~\cite{nieves}, the contribution of transitions to 1p1h final states\textemdash  described using the
local Fermi gas, or LFG, approximation\footnote{In the LFG model the nucleon Fermi momenta of isospin-symmetric nuclei depend on position through $k_F(r) = [3 \pi^2 \varrho_A(r)]^{1/3}$, $\varrho_A(r)$ being the density distribution normalised to the nuclear charge Z = A/2.}\textemdash is supplemented with those arising from processes involving MEC and
long-range RPA correlations, obtained within the diagrammatic scheme originally proposed in Ref.~\cite{RPA2}.

The hybrid model of Ref.~\cite{SuSa} combines the results of the phenomenological scaling analysis of electron scattering
data \cite{yscaling}
with a theoretical calculation of MEC contributions carried out
within the RFGM.

Overall, Fig.~\ref{degeneracy} shows that,
up to a 10\% normalization uncertainty~\cite{nieves}, the two models yield comparable
descriptions of the data. However, their predictions result from the combination of different reaction mechanisms.

Panel (A) indicates that, according to the model of Ref.~\cite{nieves}, the calculation including 1p1h final states only 
(dashed line labelled 1p1h) provides
a good approximation to the full result, represented by the solid line. The corrections arising from MEC (dot-dash line)
and long range correlations (dotted line labelled RPA) turn out to largely cancel one another. On the other hand, panel (B) suggests that, once
single-nucleon knock out processes (dashed line labelled QE) and MEC (dot-dash line) are taken into account, the addition of long-range correlations
is not needed to explain the data. 

\section{Single-nucleon emission}
\label{singlenucleon}

The picture emerging from Fig.~\ref{degeneracy} clearly calls for a deeper analysis. As a first step, it is very important to realize that
studying the role of mechanisms more complex than
the excitation of 1p1h final states is only useful to the extent to which the 1p1h sector, providing the dominant contribution to the cross 
section over a broad range of neutrino energies, see Fig.~\ref{complexity}, is fully under control.
In this context, the results shown in Fig.~\ref{degeneracy} do not appear to be very useful.

The 1p1h contribution of Ref.~\cite{nieves} [dashed line of panel (A)] is obtained within the independent particle picture of
the nucleus, according to which all single-nucleon levels belonging to the Fermi sea are filled with unit probability.
However, the data collected over fifty years of $(e,e^\prime p)$ experiments~\cite{mougey:review,benhar:npn} have unambiguously demonstrated that the occupation probability of shell model states
is in fact sizably reduced\textemdash by as much as $\sim30-35 \%$ in the case of valence states\textemdash  by correlation effects.

The dashed line labelled QE in panel (B) also fails to provide an accurate estimate of the cross section in the 1p1h sector, because the empirical scaling function
 includes additional contributions from processes involving the excitation of 2p2h final states, driven by ground state correlations, which are known to be non negligible~\cite{BLR}.

Accurate and detailed information on single nucleon knock out processes leading to the excitation of 1p1h final states has been obtained
studying the reactions
\begin{equation}
e + A \to e^\prime + p + (A-1)_{\rm B} \ ,
\label{eep}
\end{equation}
in which the scattered electron and the outgoing proton
are detected in coincidence, and the recoiling nucleus is left in a bound state.
In the absence of final state interactions (FSI), the effects of which can be taken into account as corrections, the $(e,e^\prime p)$ cross
section reduces to the simple factorised form
\begin{equation}
d \sigma_{\rm eA} = \frac{|{\bf p}|}{ T_{p} + m} P_h(p_m, E_m) \ d \sigma_{ep} \ ,
\label{eep:xsec}
\end{equation}
with the missing momentum and missing energy defined in terms of {\em measured} kinematical quantities as
\begin{equation}
\label{eep:kin}
p_m = | {\bf p} - {\bf q} |  \ \ \ , \ \ \ E_m = \omega - T_{p} - T_{A-1}   \ .
\end{equation}
In the above equations, $\omega$ is the energy transfer, ${\bf p}$ and $T_{p}$ denote the momentum and kinetic energy of the emitted proton, respectively,
and  $T_{A-1} = p_m^2/2M_{A-1}$  is the kinetic energy of the residual nucleus of mass $M_{A-1}$. Finally, the elementary cross section 
$d \sigma_{ep}$ provides a fully relativistic description of the electromagnetic interaction with a bound moving nucleon~\cite{defo}.

Equation~(\ref{eep:xsec}), that can be seen as a straightforward implementation of the factorisation {\em ansatz}, shows that a measurement of the $(e,e^\prime p)$  cross section
gives access to the hole spectral function $P_h(p_m, E_m)$, describing the probability to remove a nucleon of momentum
$p_m$ from the target nucleus, leaving the residual system with excitation energy $E_m$. 
Being trivially related to the  imaginary part of the two-point Green's function, the spectral function admits an {\em exact} decomposition
into pole and continuum contributions~\cite{BFFZ}, known as K\"all\'en-Lehman representation. This analysis allows a model independent
identification of single-nucleon emission processes, such as those of Eq.~(\ref{eep}), associated with 1p1h final states. From the experimental point of view,
these reactions are signaled by the presence of sharp spectroscopic lines in the missing energy spectra measured at low to moderate
$p_m$ and $E_m$, typically $p_m \textless  300 \ {\rm MeV}$ and $E_m \textless 30 \ {\rm MeV}$.

\begin{figure}[h!]
\begin{center}
\resizebox{0.5\columnwidth}{!}{\includegraphics{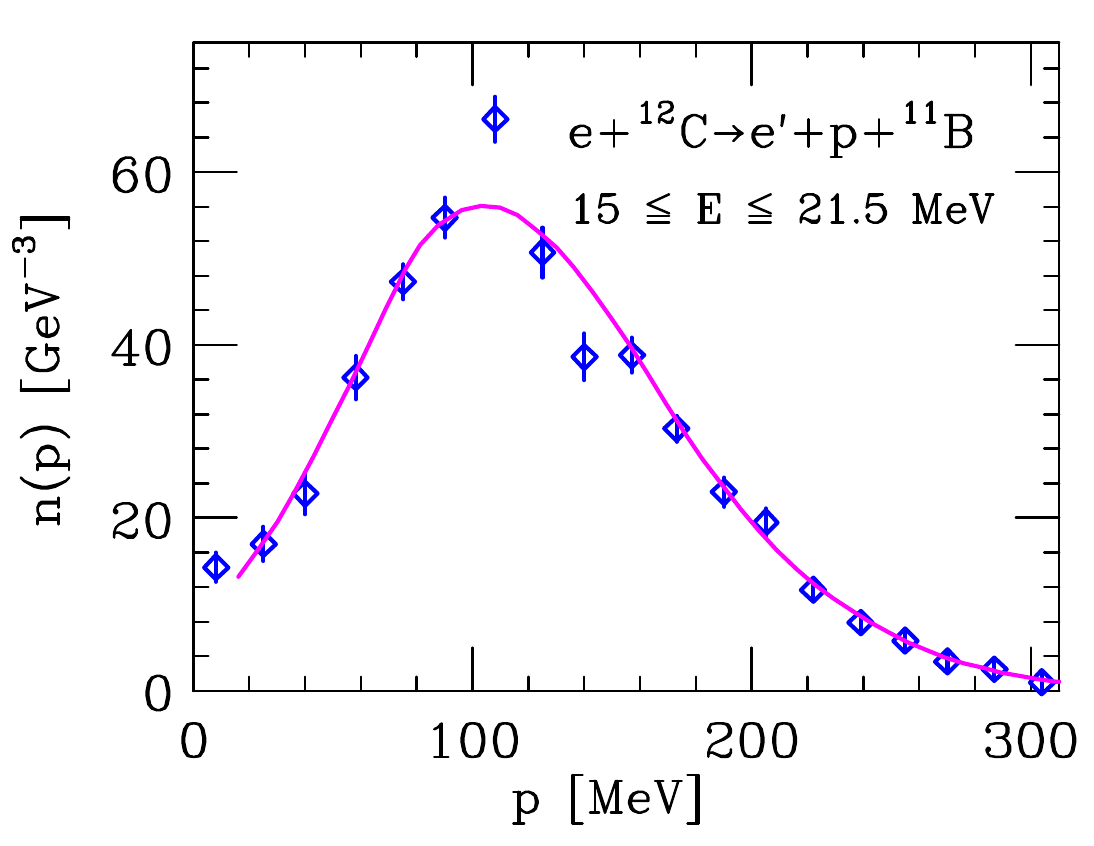} }
\end{center}
\caption{Momentum distribution of the valence $P$-states of carbon, obtained from the $(e,e^\prime p)$ cross section
of Ref.~\cite{mougey:76}. The solid line shows the momentum distribution obtained from the spectral function of
Ref.~\cite{LDA}, corrected to take into account the effects of final state interactions (FSI).}
\label{momdis}
\end{figure}

Proton knock out from carbon in the kinematical region corresponding to single-nucleon emission has been  thoroughly investigated by
Mougey {\em et al.} in the 1970s, using the
electron beam delivered by the Accelerateur Lineaire de Saclay (ALS) \cite{mougey:76}. The momentum distributions of the shell model states with quantum numbers specified by the index $\alpha$ ($\alpha = S, P$), defined as\footnote{Here we use standard spectroscopic notation, 
according to which $S$ and $P$ states correspondo to orbital angular momentum $\ell = 0$ and 1, respectively.} 
 \begin{equation}
\label{momdis:def}
n_\alpha(p_m) =  \int_{E^\alpha_{\rm min}}^{E^\alpha_{\rm max}} P_h(p_m,E_m)  \ dE_m \ ,
\end{equation}
have been obtained using the spectral function extracted from the
${^{12}{\rm C}}(e,e^\prime p){^{11}{\rm B}}$ cross section, with the integration regions being chosen
in such a way as to include the corresponding spectroscopic lines.
As an example, Fig.~\ref{momdis} shows the momentum distribution of the valence P-states,
computed  using Eq.~(\ref{momdis:def}) with  $E^P_{\rm min}~= ~15 \ {\rm MeV}$ and $E^P_{\rm max}~=~22.5 \ {\rm MeV}$
~\cite{mougey:76}. Integration over $p_m$ yields the spectroscopic factor,  providing a measure of the occupation
probability. The resulting value, $Z_P = 0.625$, implies that dynamical effects not taken into account within the
independent particle model reduce the average number of P-state protons from 4 to 2.5.

The data of Ref.~\cite{mougey:76}
have been combined with the results of accurate theoretical calculation of the continuum component of the spectral function of isospin symmetric nuclear matter~\cite{BFF}
to obtain the full carbon spectral function within the Local Density Approximation (LDA)~\cite{LDA}.

The P-state momentum  distribution computed from Eq.~(\ref{momdis:def}) using the spectral function of Ref.~\cite{LDA}, corrected for FSI following the procedure discussed in Ref.~\cite{mougey:76}, is shown by the solid line of Fig.~\ref{momdis}. It clearly appears that both shape and normalization are accurately accounted for. The spectroscopic
factor, $Z_P = 0.64$, turns out to be within $\sim2\%$ of the experimental value.

\begin{figure}[h!]
\vspace*{.15in}
\begin{center}
\resizebox{0.5\columnwidth}{!}{\includegraphics{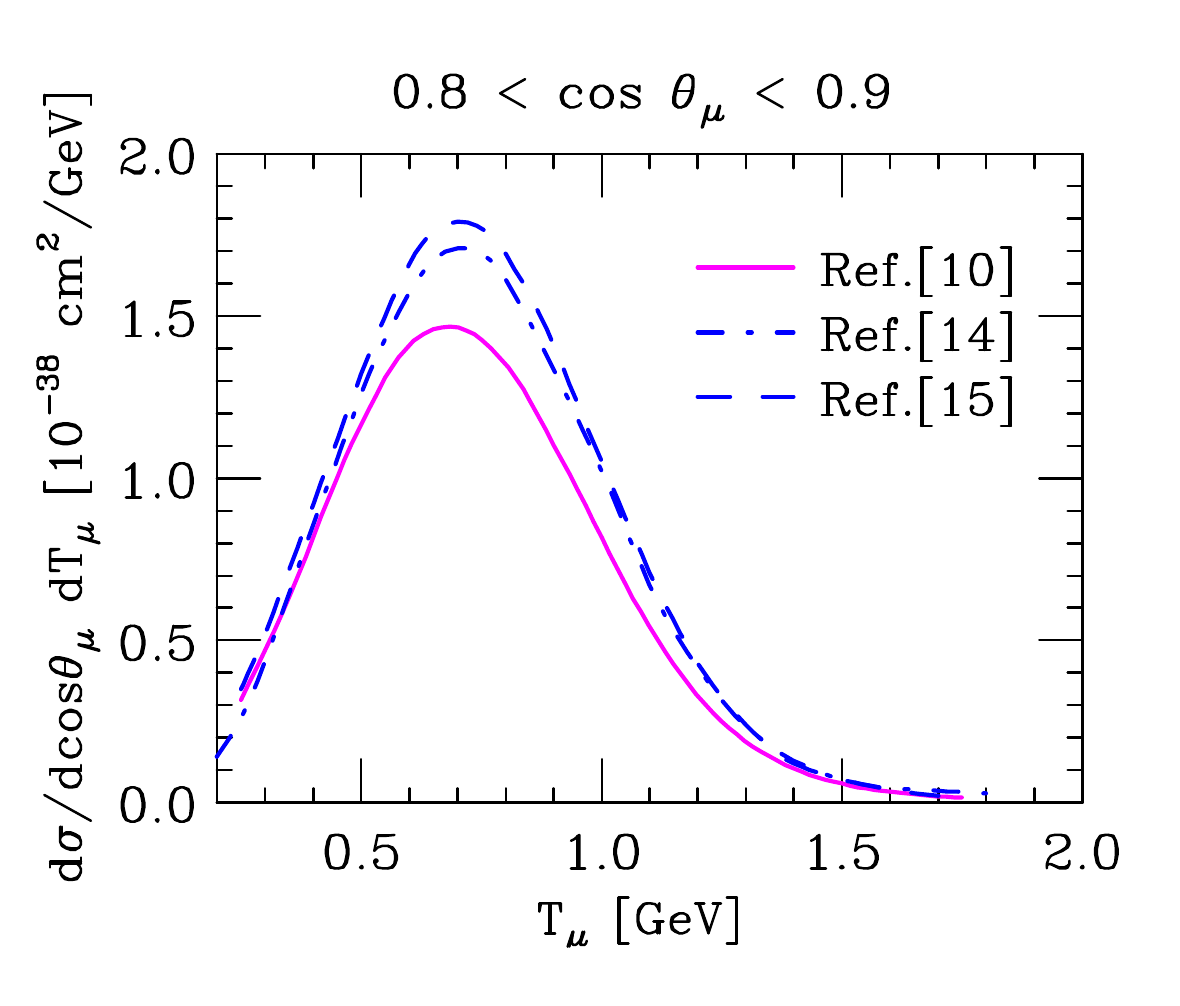} }
\end{center}
\caption{Double differential $\nu_\mu$-carbon cross section in the QE channel, averaged over the MiniBooNe neutrino flux. The dashed and solid lines
show the single nucleon emission contributions\textemdash corresponding to 1p1h final states\textemdash
obtained from the model of Nieves {\em et al.}~\cite{nieves} and the spectral function formalism of Ref.~\cite{coletti},
respectively. The dot-dash line represents the quasi elastic cross section obtained from the scaling analysis of Ref.~\cite{SuSa}. }
\label{degeneracy2}
\end{figure}

A comparison with the data of Refs.~\cite{rohe1,transparency,rohe2}, reporting the results of a measurement of the $(e,e^\prime p)$ cross section at large $p_m$ and $E_m$ performed at the Thomas Jeffetson National
Accelerator Facility (JLab), shows that
the spectral function of Ref.~\cite{LDA} also provides a quantitative description of the contribution arising from nucleon-nucleon correlations. The continuum strength integrated over
the region covered by the JLab experiment turns out to be $0.61\pm0.06$, to be compared with the theoretical value 0.64~\cite{rohe2}.

The pole component of the spectral function of Ref.~\cite{LDA} can be employed to obtain the 1p1h contribution to the flux integrated CCQE $\nu_\mu$-carbon cross section within the implementation of the factorisation scheme
described in Refs.~\cite{benhar_PRD,coletti,ankowski}, in which FSI effects are also taken into acount. In Fig.~\ref{degeneracy2} the results of this calculation are compared with the 1p1h cross section of Ref.~\cite{nieves} and the QE result of Ref.~\cite{SuSa}.
As it was to be expected on the basis of the discussion of Section~\ref{sec2}, both the local Fermi gas model and the phenomenological scaling analysis significantly overpredict  the 1p1h cross section
obtained using the spectral function of Ref.~\cite{LDA}, which is strongly constrained by $(e,e^\prime p)$ data. Note that the $\sim20\%$ difference at the peak is about the same size as the discrepancy between the MiniBooNE data
and the results of Monte Carlo simulations based on the RFGM, that stirred a great deal of debate on the need to use an {\em effective axial mass} in modeling neutrino-nucleus interactions~\cite{BM:PRD}.

\section{Two-nucleon emission}
\label{extended}

As pointed out above, two-nucleon emission can be triggered by nucleon-nucleon correlations in the initial state, FSI, and interactions 
involving two-nucleon currents. Within the factorisation scheme, correlations can be taken into account by the use of realistic hole spectral functions, 
including the continuum contribution extending to large momentum and energy~\cite{BFFZ}. In principle, the spectral function formalism provides 
a consistent framework for the description of both initial state correlations and FSI~\cite{omar:FSI}. However, in the kinematical region of large momentum 
transfer the motion of the struck nucleon in the final state cannot be described in the non relativistic approximation. In the well established convolution 
approach~\cite{omar:FSI}, widely employed to study electron-nucleus scattering data~\cite{ankowski}, the inclusive cross section at fixed beam energy 
is written in the form
\beq
\frac{d\sigma}{d\omega d\Omega} = \int d\omega^\prime f_q(\omega - \omega^\prime) \frac{d\sigma_{0}}{d\omega^\prime d\Omega} \ , 
\eeq
where $\omega$ and $q$ are the energy and momentum transfer, $\Omega$ is the solid angle specifying the direction of 
the emitted lepton, and $d\sigma_0$ denotes the cross section in the absence of FSI. The derivation of the folding function embodying
the effects of FSI is discussed in Refs.~\cite{ankowski,omar:FSI}.  

The approach based on factorisation of the nuclear cross section and the spectral function formalism has been generalised to the description of
transition matrix elements involving MEC and 2p2h final states~\cite{BLR,RLB}. In processes driven by two-nucleon currents, the relevant nuclear amplitudes are those contributing to the two-nucleon spectral function, whose derivation for the case of isospin-symmetric matter is is discussed in 
Ref.~\cite{pke2}. 

\begin{figure}[h!]
\vspace*{.15in}
\begin{center}
\resizebox{1.00\columnwidth}{!}{\includegraphics{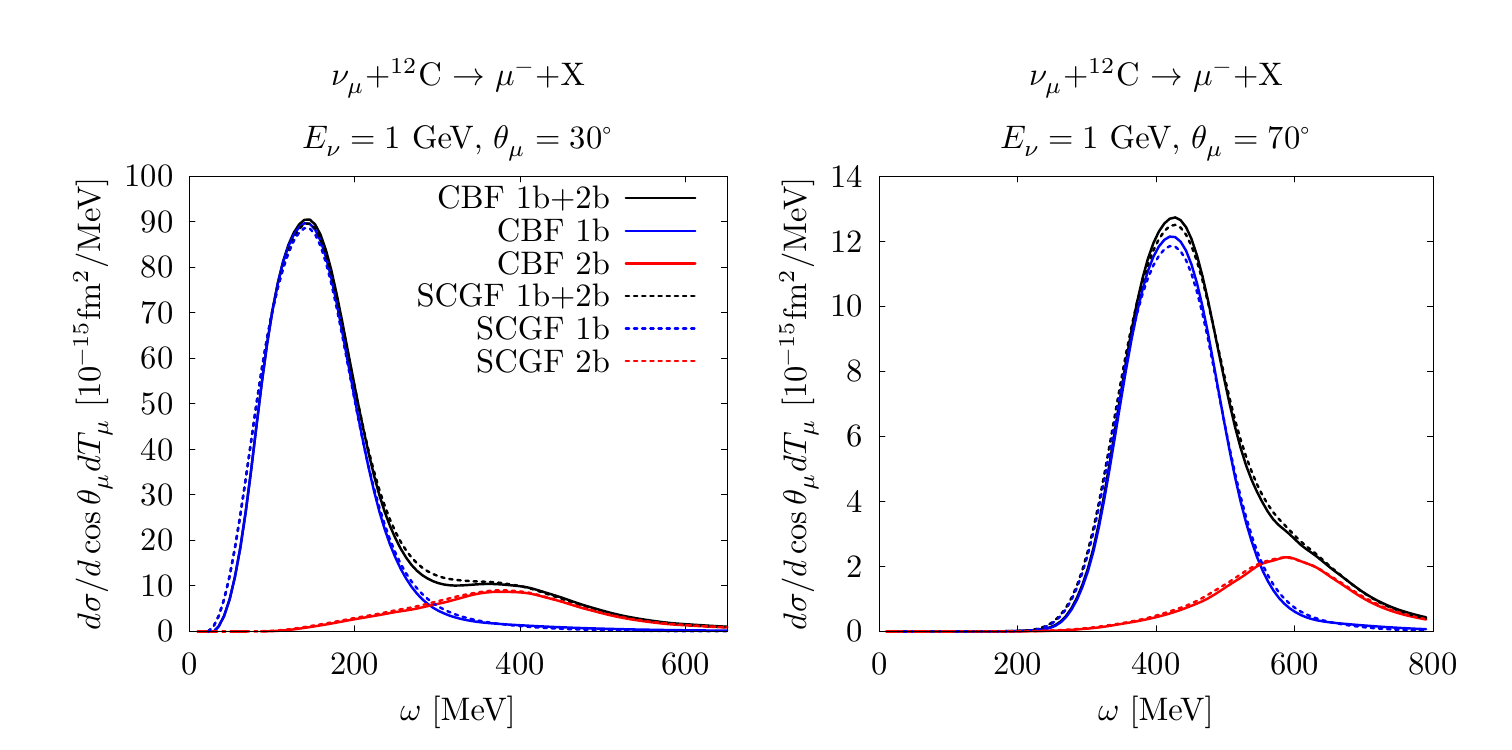} }
\end{center}
\caption{Left: double differential cross section of the process $\nu_\mu + {^{12}{\rm C}} \to \mu^- + X$ at beam energy 1 GeV and muon 
emission angle 30 deg. The lines labelled $1b$ ($2b)$ have been obtained including one-body (two-body) current contributions only, while the 
line labelled $1b + 2b$ correspomds the full result. The labels CBF and  SCGF refer to the many-body technique employed in the calculation of
nuclear amplitudes. RIght: same as in the left panel, but for beam energy 1 GeV and muon emission angle 70 deg.}
\label{MEC}
\end{figure}

Figure~\ref{MEC} illustrates the contributions associated with one- and two-body currents, denoted $1b$ and $2b$, to the $\nu_\mu$-carbon 
differential cross section, displayed as a function of energy transfer $\omega$ for two different kinematical setups. 
It clearly appears that the MEC contribution to the cross section is sizeable, and peaked at $\omega$ larger than that corresponding to 
the one-body current contribution. In addition, 
a comparison between the curves labelled CBF and SCGF\textemdash obtained using Corelated Basis Function perturbation theory~\cite{BFF} and 
the Self Consistent Green's Function approach~\cite{SCGF}, respectively\textemdash  shows that the theoretical uncertainty arising from the use of different many-body techniques to perform the calculations of nuclear amplitudes is small, and hardly visible. 

\section{Resonance production and deep inelastic scattering}
\label{inelastic}

Within the factorisation scheme the response tensor of Eq.~(\ref{dsigmaA:2}) reduces to the form\footnote{Note that this expression 
applies to isospin-symmetric targets, in which the neutron and proton spectral functions can be assumed to be identical.}
 \beq
W_A^{\alpha \beta}  = \int d^3p dE P_h(p,E)  \left[ Z W_p^{\alpha \beta} + N W_n^{\alpha \beta}  \right]  \ ,
\eeq
where  $W_N^{\alpha \beta}$, with $N = p$ or $n$, is the tensor describing the interactions of a single nucleon of momentum 
${\bf p}$ and removal energy  $E$, 
while $Z$ and $N$ denote the target charge and number of neutrons, respectively. Based on its Lorentz transformation properties, the nucleon 
tensor $W_N^{\alpha\beta}$ can be written in terms of five structure functions according to
\bea
W_N^{\alpha\beta}&=&-g^{\alpha\beta}\,W_1^N+\widetilde p^\alpha\,\widetilde p^\beta\,\frac{W_2^N}{m_N^2}+
i\,\varepsilon^{\alpha\beta\rho\sigma}\,\widetilde q_{\rho}\,\widetilde p_\sigma\,
\frac{W_3^N}{m_N^2} \\ 
\nonumber
&+& \widetilde q^\alpha\,\widetilde q^\beta\,\frac{W_4^N}{m_N^2} 
+ (\widetilde p^\mu\,\widetilde q^\nu+\widetilde p^\nu\,\widetilde q^\mu)\,\frac{W_5^N}{m_N^2} \  , 
\eea
with $\widetilde p \equiv( \sqrt{{\bf p}^2 + m_N^2},{\bf p})$, 
$\widetilde q\equiv(\widetilde \omega,{\bf q})$, and $\widetilde \omega \approx \omega -E$~\cite{physrep:nu}. 

The structure functions depend on the kinematical 
variables through the independent scalars $\widetilde p^2$, $\widetilde q^2$, and $(\widetilde p \cdot \widetilde q)$. 
In QE processes, defined by the additional condition $W^2~=~(\widetilde p + \widetilde q)^2 = m_N^2$, they can be written as
\beq
W_i^N = \widetilde W_i^N \ \delta(\widetilde \omega + \frac{\widetilde q^2}{2 m_N}) \ ,
\eeq
and the $W_i^N$ are defined in terms of the vector and axial-vector form factors of the nucleon. 

Conceptually, the generalisation of the factorisation approach to describe resonance production, driven by elementary processes such as
\begin{equation}
\label{delta++}
\nu_\mu + p \to \mu^- + \Delta^{++}  \to \mu^- + p + \pi^+ \ ,
\end{equation}
where $\Delta^{++}$ denotes the $P_{33}(1232)$ nucleon resonance, only requires minor changes \cite{Benhar:2006nr}.
In this case, the neutrino-nucleon cross section involves  the matrix elements of the weak current describing nucleon-resonance transitions.
As a consequence, the structure functions\textemdash which can still be written in terms of phenomenological vector and axial-vector form factors\textemdash  depend 
on both $\widetilde Q^2= -\widetilde q^2$ and $W^2$, the squared invariant mass of the state of the hadronic state produced at the weak interaction vertex. In addition, the energy conserving $\delta$-function 
is replaced by a Breit-Wigner function, accounting for the finite width of the resonance. 

The authors of Ref.~\cite{VBM_inelastic} have carried out a consistent calculation of QE and inelastic neutrino-carbon interactions based on factorisation, 
using the spectral function of Ref.~\cite{LDA}.
Besides the prominent $P_{33}(1232)$ state, providing the largest contribution to the cross section, they have taken into account the three isospin $1/2$ 
states\textemdash  $D_{13}(1520)$, $P_{11}(1440)$, and $S_{11}(1535)$\textemdash comprised in the so-called second resonance region. 
The numerical results have been obtained using the parametrization of the structure functions described  in Refs~\cite{res1,res3,res2}. Within this approach, the vector form factors
are constrained by electroproduction data, while the axial couplings are extracted from the measured resonance decay rates, exploiting the Partially Conserved 
Axial Current (PCAC) hypothesis.   

As mentioned above, from the observational point of view DIS is associated with hadronic final states comprising more than one pion.
In principle, the three nucleon structure functions determining the neutrino-nucleon cross section in the DIS  regime\textemdash $W_1$, $W_2$ and $W_3$\textemdash 
may be obtained combining measured neutrino and antineutrino scattering cross sections. However, because the available structure functions have been
extracted from {\em nuclear} cross sections (see, e.g., Ref.~\cite{CDHS}),
their use in {\em ab initio}  theoretical studies, aimed at identifying nuclear effects, entails obvious conceptual difficulties.

An alternative approach, allowing to obtain the structure functions describing DIS on isolated nucleons, can be developed within the conceptual
framework of the quark-parton model, exploiting the large database of accurate DIS data collected using charged lepton beams and  hydrogen and deuteron targets, see, e.g., Ref.~\cite{roberts}.
Within this scheme,  the function $F_2^{\nu N} = \omega W_2$, where $W_2$ is the structure function of  an isoscalar nucleon,
can be simply related to the corresponding structure function extracted from electron scattering data, $F_2^{e N}$ through
\begin{equation}
F_2^{\nu N}(Q^2,x) = \frac{18}{5} \  F_2^{e N}(Q^2,x)  \ ,
 \label{DIS1}
\end{equation}
where $x=Q^2/2m \omega$ is the Bjorken scaling variable. In addition, the relation\footnote{For the sake of simplicity, here, and in what follows, the contributions of $s$ and $c$ quarks is not taken into account.}
\begin{equation}
\label{DIS2}
x F_3^{\nu N}(Q^2,x)  = x \ [ \ u_{\rm v}(Q^2,x)  + d_{\rm v}(Q^2,x)  \ ] \ ,
\end{equation}
where $F_3^{\nu N} = \omega W_3$ and  
$u_{\rm v}$  and $d_{\rm v}$ denote the valence quark distributions, implies
\begin{equation}
\label{DIS3}
x F_3^{\nu N}(Q^2,x)  = F_2^{eN}(Q^2,x) 
- 2 x \  [ \overline{u}(Q^2,x) + \overline{d}(Q^2,x)]   \ .
\end{equation}
Using Eqs.~(\ref{DIS1})-(\ref{DIS3}) and the Callan-Gross relation~\cite{roberts},  linking  $F_1^{\nu N} =  m_N W_1$ to $F_2^{\nu N}$,  one can readily obtain all the relevant weak
structure functions from the existing parametrizations of the measured electromagnetic structure function and the antiquark distributions $\overline{u}$ and $\overline{d}$, see, e.g., Ref~\cite{GRV98}.
Alternatively, the quark and antiquark distributions can be also used to obtain the structure function $F_2^{e N}$ from
\begin{equation}
F_2^{e N}(Q^2,x)  = x \ \frac{5}{18}  [ \   u(Q^2, x)  + \overline{u}(Q^2,x)  \\
  + d(Q^2,x) + \overline{d}(Q^2,x) \ ]  \ .
\label{DIS0}
\end{equation}
The authors of Ref.~\cite{VBM_inelastic} have used Eqs.~(\ref{DIS1}), (\ref{DIS2}) and  (\ref{DIS0}) with the parton distributions of Ref.~\cite{GRV98}, which are available for $Q^2~\geq~Q^2_{\rm min}~=~0.8$~GeV$^2$. At lower values of $Q^2$, 
the parton distributions have been assumed to be the same as at $Q^2=Q^2_{\rm min}$. 

Note that the above procedure rests on the tenet, underlying the factorisation scheme, that the elementary neutrino-nucleon interaction is {\em not} affected by the
presence of the nuclear medium, the effects of which are accounted for with the substitution $\omega \to \widetilde \omega$~\cite{physrep:nu}. While this assumption is strongly supported by electron-nucleus scattering data in the quasi elastic channel, showing no evidence of medium modifications of the nucleon vector form factors, it has to be mentioned that analyses of neutrino DIS data are often
carried out within a conceptually different approach, allowing for medium modifications of either the nucleon structure functions \cite{petti,haider}, or of the parton distributions entering their definitions \cite{kumano}.

\begin{figure}[h!]
\vspace*{-.15in}
\begin{center}
\resizebox{0.485\columnwidth}{!}{\includegraphics{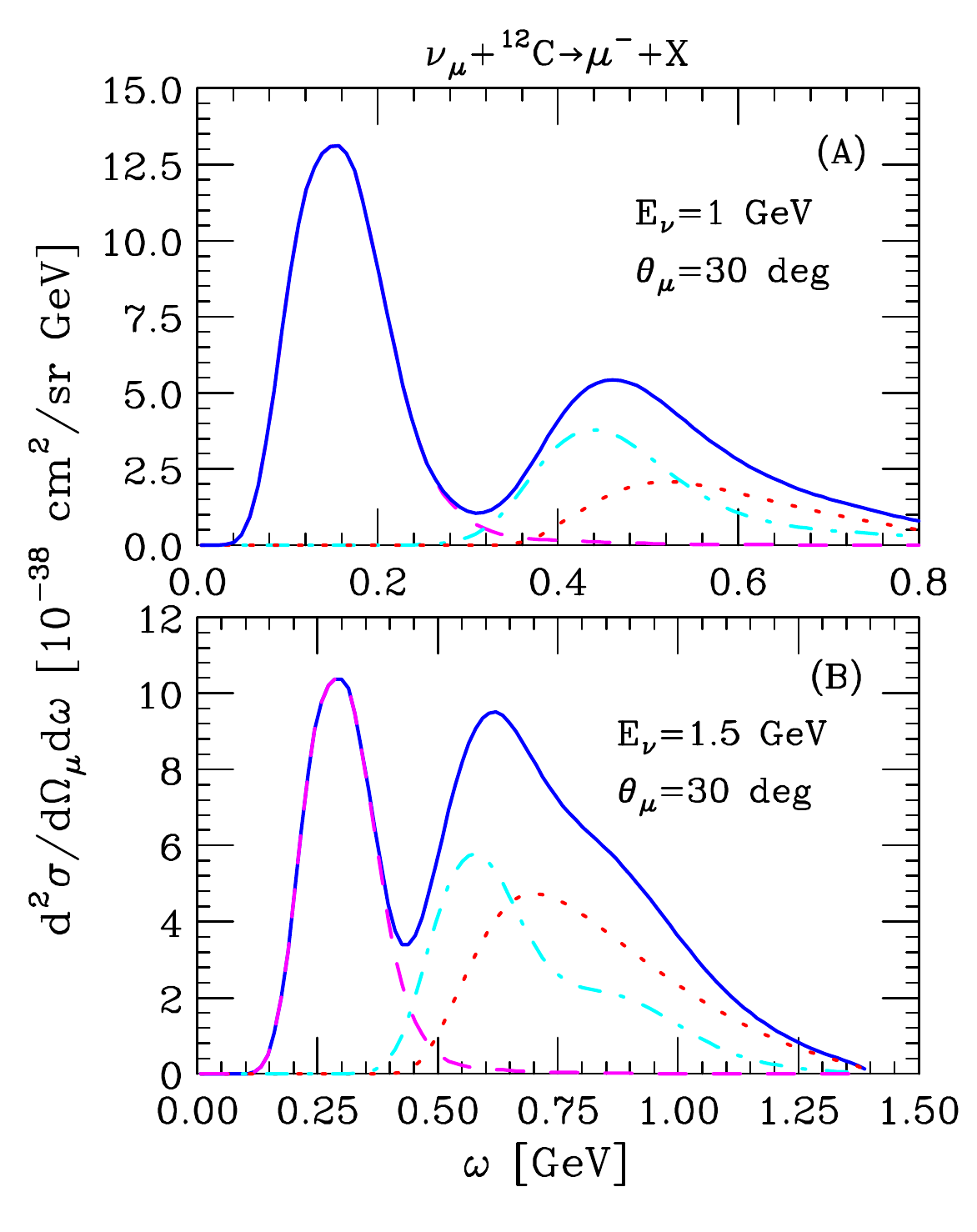} }
\resizebox{0.475\columnwidth}{!}{\includegraphics{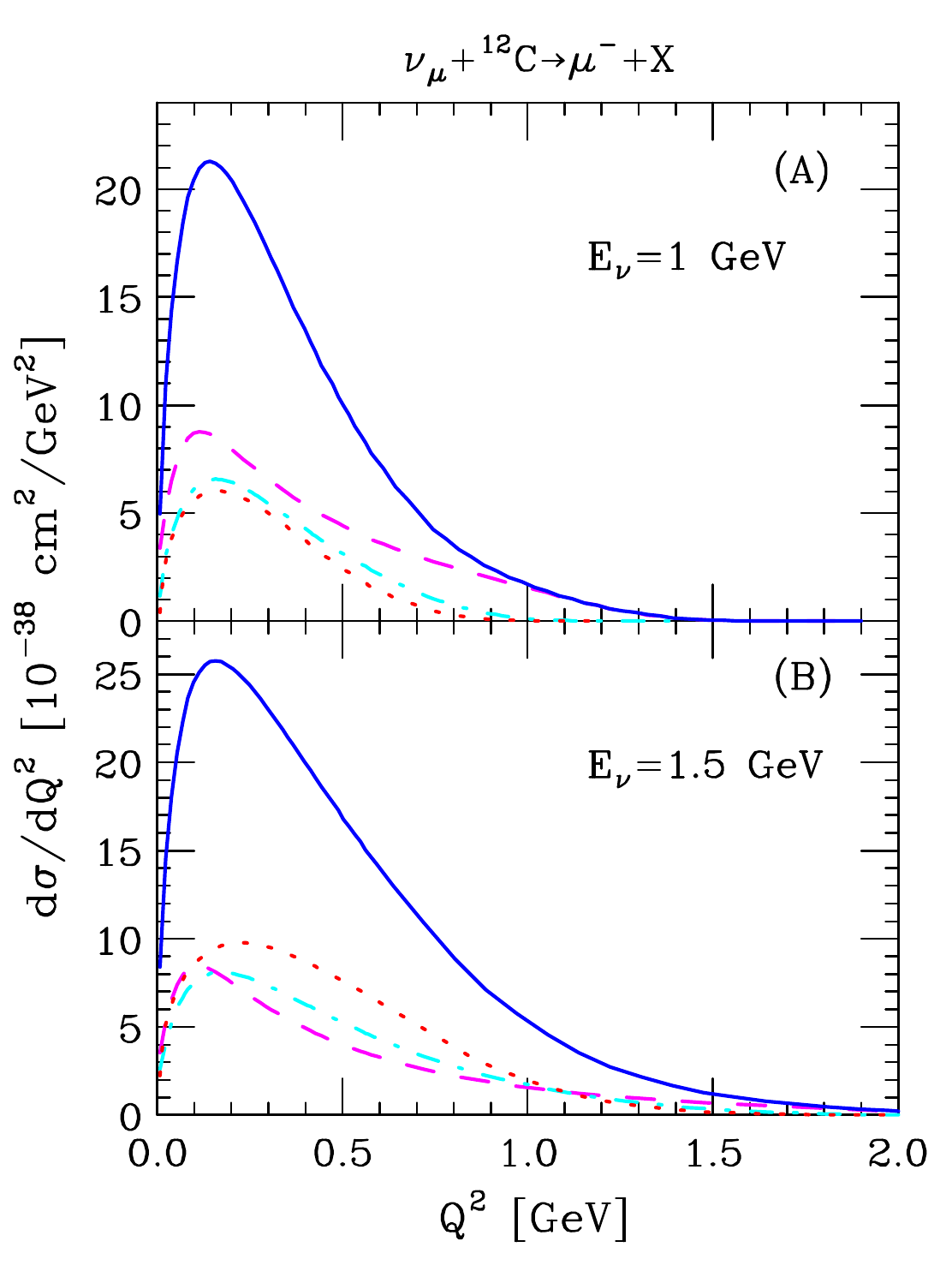} }
\end{center}
\caption{Left: double-differential cross section of the  scattering process $\nu_\mu + {^{12}C} \to \mu^- + X$ at fixed muon emission angle $\theta_\mu = 30 \ {\rm deg}$, and beam energies 
$E_\nu = $ 1 GeV (A) and 1.5 GeV (B), displayed as a function of $\omega = E_\nu - E_\mu$. The dashed, dot-dash and dotted lines 
correspond to QE scattering, resonance production and DIS, respectively. The sum of the three contributions is represented 
by the full line.
Right: $Q^2$-distribution of the process $\nu_\mu + {^{12}C} \to \mu^- + X$ at fixed neutrino energy $E_\nu$ = 1 Gev (A) and 1.5 GeV (B). The meaning of the lines is the 
same as in the left panel.}
\label{d2sigma}
\end{figure}

The results of calculations of the electron-nucleus cross sections have provided ample evidence that the 
approach based on factorisation and the spectral function formalism, {\em involving no adjustable parameters}, is capable to deliver a quantitative description of the double-differential electron-nucleus cross sections\textemdash measured at fixed beam energy and electron scattering angle\textemdash
in both the quasi elastic and inelastic sectors~\cite{LDA,electron,benpan}. The left panel of Fig.~\ref{d2sigma} shows the results of the extension of these analyses
to the case of  neutrino-carbon interactions reported in Ref.~\cite{VBM_inelastic}. The calculations have been carried out using the spectral function of Ref.~\cite{LDA} and
setting the muon emission angle to $\theta_\mu = 30 \ {\rm deg}$. Comparison between panels (A) and (B), corresponding to $E_\nu =$ 1 and 1.5 GeV, respectively, illustrates 
how the relative weight of the different reaction mechanisms changes with increasing neutrino energy.

\begin{figure}[t!]
\begin{center}
\resizebox{0.6\columnwidth}{!}{\includegraphics{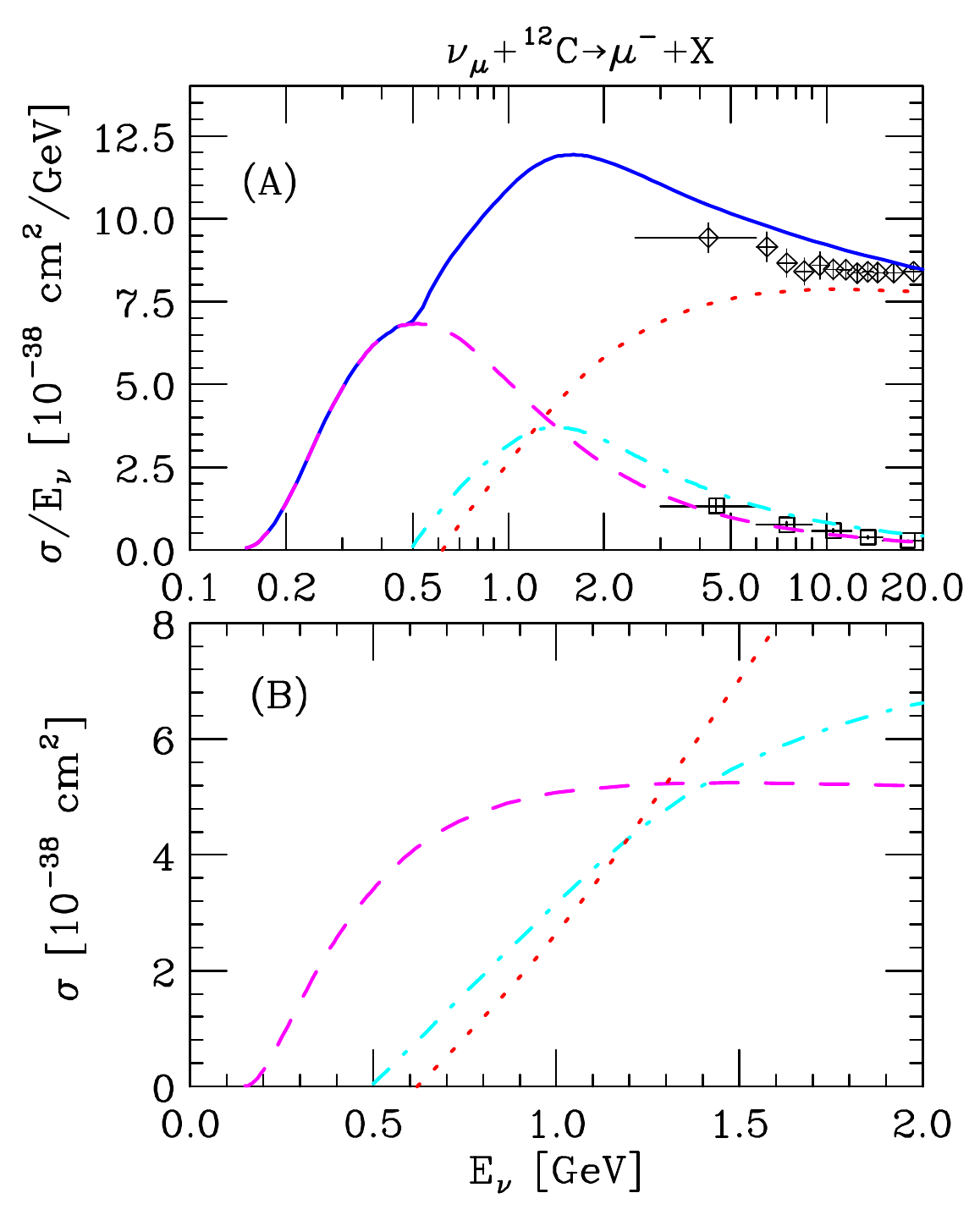} }
\end{center}
\caption{Total cross section of the reaction $\nu_\mu + {^{12}C} \to \mu^- + X$ as a function of neutrino energy. The dashed, dot-dash and dotted lines of panel (A) 
represent the contributions of QE, resonance production and DIS processes. Panel (B) shows the $E_\nu$-dependence of the ratio $\sigma/E_\nu$.
The meaning of the dashed, dot-dash and dotted lines is the same as in panel (A). The full line corresponds to the sum of the three contributions.
Diamonds and squares represent the data of Refs.~\cite{NOMAD_tot} and \cite{NOMAD_CCQE}, respectively.}
\label{sigmatot}
\end{figure}

The $Q^2$-distributions, obtained from the double-differential cross sections by integrating over $\cos \theta_\mu$, are displayed in the rigt panel of Fig.~\ref{d2sigma}.
At both $E_\nu =$ 1 and 1.5~GeV, the full $d\sigma/dQ^2$, corresponding to the solid line, exhibits a pronounced maximum at $Q^2\lsim 0.2 \ {\rm GeV}^2$. 
Further integration over $Q^2$ yields the total cross section, $\sigma$, whose behavior as a function of the neutrino energy $E_\nu$ is illustrated in Fig.~\ref{sigmatot}. 
Panels (B) and (A) show $\sigma$ and the ratio $\sigma/E_\nu$, respectively, as well as
the contributions corresponding to the QE, resonance production, and DIS channels. It is apparent that, while at $E_\nu \lsim 0.8$ GeV QE interactions dominate, 
the inelastic cross section rapidly increases with energy. At  $E_\nu \approx 1.3$~GeV, the contributions arising from the three reaction channels turn out to be about the same. 

For comparison, in panel (B) we also report, as diamonds, the $\nu_\mu$-carbon total cross section measured by the NOMAD collaboration \cite{NOMAD_tot}.
It turns out that, while the energy-dependence of the data at $E_\nu \gsim 10$ GeV is well reproduced by the theoretical prediction of the DIS contribution, represented by the dotted line,  
the results of the full calculation, corresponding to the solid line, sizably exceed the measured cross section. In view of the fact that the QE cross section obtained from the NOMAD 
data of Ref.~\cite{NOMAD_CCQE}, shown by the open squares, turns out to be in close agreement with the results of theoretical calculations, 
this discrepancy is likely to be ascribed to double counting between resonance production and DIS contributions, which are very hard to identify in a truly model independent fashion.  In principle, this problem may be circumvented using structure functions obtained from a global fit of proton and deuteron data, including 
both resonance production and DIS. Unfortunately, however, such an analysis of neutrino interactions is not available. Furthermore, as pointed out by the authors of Ref.~\cite{benmel}, the existing parametrizations of the electromagnetic structure functions~\cite{BR_fit} fail to provide an accurate description 
 of the inelastic electron-deuteron cross section at $Q^2\lsim 1$ GeV$^2$.

\section{Summary and outlook}
\label{sec4}

The description and interpretation of the neutrino-nucleus cross section\textemdash the quantitative understanding of which is a needed requisite for
oscillation analyses\textemdash involve serious conceptual and computational issues, primarily due to the variety of reaction mechanisms
providing significant contributions to the flux-averaged signals.  This issue was recognised and discussed over a decade ago by the authors of Refs.~\cite{coletti,neutrino:2010}. In these papers,  it is argued that the 
challenges implied in the theoretical description of flux-integrated cross sections called for the development of a {\em new paradigm}, suitable for 
the derivation of a {\em consistent} theoretical approach applicable in the broad kinematical range relevant to accelerator-based neutrino experiments. 
In view of the recent measurements of inclusive double differential cross sections, and 
of the pioneering analyses of the Transverse Kinematics Imbalance (TKI) in $0\pi$ events, such an approach appears to be all the more indispensable today.

The theoretical description of the data requires a unified model of neutrino-nucleus interactions\textemdash  
applicable to a variety of nuclear targets, reaction channels and kinematical regimes\textemdash the accuracy of which can only be assessed testing the ability to reproduce
independent data sets, most notably electron scattering data.

In spite of many remarkable progresses, the available theoretical models, 
 while yielding
a fairly good description of the QE cross sections, fail to provide an unambiguous interpretation of the underlying mechanisms. 
In order to establish if, and to what extent, the agreement between theory and experiments is, in fact, accidental, the near-degeneracy between approaches based on different assumptions must be resolved.
The measured cross sections of {\em exclusive} $(e,e^\prime p)$  processes\textemdash the study of which allows to isolate the contribution
of single nucleon knock out processes, leading to the excitation of 1p1h final states\textemdash  provide the ideal tool for gauging the ability
of different models to describe the dominant reaction mechanism in the QE, or $0\pi$, sector. 
In order to acquire additional information, needed for the interpretation of signals detected using the liquid argon technology,  the
available dataset is being augmented with the ${\rm Ar}(e,e^\prime p)$ and ${\rm Ti}(e,e^\prime p)$ cross sections recently measured
at Jlab, the analysis of which is underway~\cite{JLab1,JLab2}.

Factorisation of the nuclear cross section, which draws its justification from the assumptions underlying the impulse 
approximation, is a natural option to circumvent the problems associated with the description of neutrino interactions. 
As a matter of fact, besides the models based on the mean-field approximation, which are inherently factorised, 
many existing approaches exploit some level of factorisation: from the one exploiting spectral functions 
to the superscaling approach and  the short-time approximation of Quantum Monte Carlo~\cite{saori}.

The results discussed in this paper suggest that the approach based on factorisation and the spectral function formalism\textemdash
which has been extensively applied in studies of electron-nucleus cross sections\textemdash is attaining the level required for a consistent 
and systematic analysis of the flux-integrated neutrino-nucleus cross sections, in both the elastic and inelastic sectors. 
From the conceptual point of view, the main outstanding issues appear to be the inclusion of collective nuclear excitations\textemdash whose impact,  or lack thereof, needs to be firmly established\textemdash and the treatment 
of FSI in exclusive processes. 

The pioneering study carried out by the authors of Ref.~\cite{BF:PLB} indicate that collective effects, 
while being obviously beyond the impulse approximation, can be described using the same formalism employed to obtain the 
spectral functions. On the other hand, the most promising approach to the treatment of FSI involves the combination of  a suitable 
spectral function\textemdash obtained within the framework of the eikonal approximation~\cite{omar:FSI}\textemdash and semiclassical 
cascade methods, suitable to describe 
the collisions between the struck nucleon and the spectators~\cite{cascade}.
In both instances, the availability of electron scattering data will be essential. Resolving these issues will likely require a great deal of theoretical and experimental effort for several years to come.


\end{document}